# Walking through a library remotely

*Why we need maps for collections and how KnoweScape can help us to make them?*

ANDREA SCHARNHORST

There is no escape from the expansion of information, so that structuring and locating meaningful knowledge becomes ever more difficult. The question of how to order our knowledge is as old as the systematic acquisition, circulation, and storage of knowledge. Classification systems have been known since ancient times. On the Internet, one finds both classifications and taxonomies designed by information professionals and folksonomies based on social tagging. Nevertheless, a user navigating through large information spaces is still confronted with a text based search interface and a list of hits as outcome. There is still an obvious gap between a physical encounter with, for example, a library's collection and browsing its content through an on-line catalogue. This paper starts from the need of digital scholarship for effective knowledge inquiry, revisits traditional ways to support knowledge ordering and information retrieval, and introduces into a newly funded research network where five different communities from all corners of the scientific landscape join forces in a quest for knowledge maps. It can be read as a manifesto for a newly funded specific research network KnoweScape. At the same time it is a general reflection about what one has to take into account when representing structure and evolution of data, information and knowledge and designing instruments to help scholars and others to navigate across the lands and oceans of knowledge.

## 1. Introduction

   Since digitization and web technologies we seem to have any information possibly needed under our fingertips. But a closer look reveals that although we might not need to go physically to a library or archive anymore, it remains a time consuming, resource eating process to compose an overview, a literature review, a syllabus, or insights into current scientific trends from bits and pieces of information scattered around. Maybe while operating in the digital age of scholarship (Borgman, 2007) our expectations concerning the speed and easiness of information processing have changed? Maybe in the transition to the digital world we also lost something comparing to our information foraging behaviour pre-Internet? This paper takes both aspects – shifts in the rhyme of scholarship and pending issues in remote access to digital resources - as departure points to reflect about the need to design visual aids, and in particular *maps as macroscopes*  (Börner, 2011) for the navigation in massive and distributed information spaces. Eventually, we present the conceptual framework and organizational setting of a European network – the COST Action KnoweScape [1], initiated to produce such knowledge maps. We argue why a broad composition of disciplinary communities involved in this endeavour is needed. We end the paper discussing dreams and visions about knowledge maps and their possible use.

## 2. Digital scholarship, the eternal research cycle and the information challenge

Changes in scholarship in the digital age have been addressed from the very beginning of digital technologies penetrating the research process and even more with the emergence of the Internet. At early conferences of the Association for Internet Research [2] one could witness a strong presence of ethnographic research into the effect of the Internet on all forms of social and cultural activities, including scholarly research. Some authors proclaimed the rise of a new, all-encompassing discipline, *cyberscience* [3] (Nentwich, 2003), others look into the emergence of new research infrastructures accompanying new scholarly practices (Edwards et al., 2013), and where others critically question the newness of virtual knowledge production at all (Wouters et al., 2013).

There is no doubt about some research practices changing profoundly, and the need to adopt science policy measures to respond to *team science* (Wuchty et al., 2007) and *big-new-data driven science* (Ryder, 2010). Practitioners regularly express this in self-reflections about changes in their disciplines. But the natural, professional place for this kind of research are *science and technology studies, quantitative studies of science, science history and science philosophy* (cf. Latour and Woolgar, 1986). All references in the paragraph above come from this area in the landscape of science. In German, one would easily put the label *Wissenschaftsforschung* on this quite heterogeneous collection of fields, which differ by concepts and methods, unified only by "science" being their object of studies. In English, the use of the notion *science of science* (Börner and Scharnhorst, 2009) is rather hampered by the restricted meaning of *science* among English native speakers.  This is not the place to review this large body of literature, and the few remarks are meant to give a reader not familiar with those *academic tribes* (Becher, 1989) some landmarks from which to start a closer reading.

The quest for *knowledge maps* has been informed by science and technology studies. The last, more recent move in this particular stream of academic reflection *about* academics concerns new alliances between digital humanities and digital libraries (Prescott, 2012). New roles for librarians as data stewards, teachers of digital and computational literacy, and curators for data, information and knowledge are designed. We discuss later how, with this, the question of *tools* (including visual once) for information seeking moves to the foreground.

Despite profound changes in scholarship, the essence of scholarship has not changed. It still starts with iterative paths around questioning, searching, reading and thinking. Christine Wong Yap once depicted the classic creative cycle in an art project, incorporating also distortions to that theoretical construct in the actual practice (Figure 1). The creative cycle in scholarship – the research cycle - is very similar. How much meandered the actual process might be, information gathering and knowledge inquiry can always be found at the beginning.

In the eternal process of knowledge creation, a balance seems to have shifted. In the pre-Internet age, searching and looking up required substantive time and efforts. Nowadays, we can access an increasing amount of information much faster. In other words, the time we need to collect the information shrinks, but the time we need to inspect and evaluate each bit of information remains the same. The problem for information foraging (Sandstrom, 1994) is not longer access to resources but time to inspect them. Our limited individual capacity to orient ourselves in a rising ocean of scientific information seems to become the bottleneck.

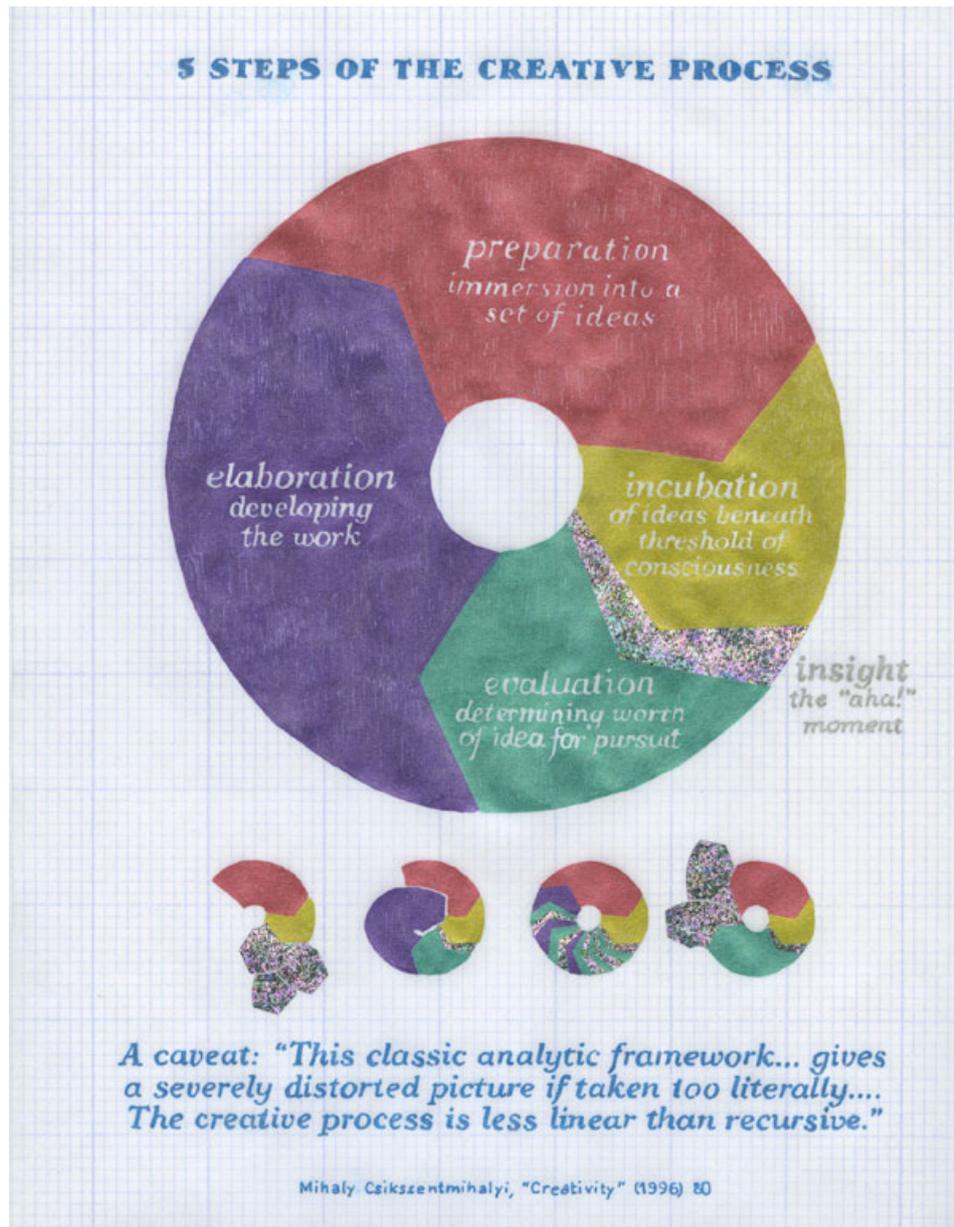

Figure 1: Christine Wong Yap, Positive Signs #1 (Five Steps of the Creative Process), 2011, glitter pen with holographic foil print on gridded vellum, 8.5 x 11 in / 21.5 x 28 cm) Source: http://blog.sfmoma.org/2011/03/positive-sign-1/

Industrialization, Big Science after World War II, and the Open Access movement in the digital age lead to a measurable exponential or hyperbolic growth of scientific literature (Price, 1963; Börner, 2010). The thread of an information flood has been articulated countlessly, often with reference to Wilson's classical quote [4] «We are drowning in information, while starving for wisdom. The world henceforth will be run by synthesizers, people able to put together the *right* information at the right time, think critically about it, and make important choices wisely.» (Wilson, 1998, p 294). However, the problem of finding the right information is by no means a new one, and probably as old as written evidence exists. It seems therefore more than appropriate to undertake a retrospection following the footsteps of historians of the information sciences.

## 3. Old and new *Delta works* to manage information (floods)

As inspiration for the quest for *knowledge maps* we would like to point to three examples of Delta Works [5] from the past, a kind of super-dikes, against the information flood: Paul Otlet's International Bibliographic Institute; Eugene Garfield's Institute for Scientific Information; and Wikipedia.

The International Bibliographic Institute was funded in 1895 (Rayward, 1997). It provided an information service where *queries*, questions, could be sent to Brussels and pointers to literature were sent back (Rayward, 2013). The answers were retrieved from the growing *Repertoire Bibliographique Universel* – according to Rayward, a database on paper, a *Europeana*-like aggregator for catalogue entries from libraries world-wide, a Google like database [6] using the Universal Decimal Classification (UDC) as indexing engine and the postal systems as network of information transmission. [7] About half a century later Eugene Garfield funded the Institute for Scientific Information in Philadelphia - again a response to an acceleration of scientific production by inventing new information services (Wouters, 1999). This time, classification and abstracts services would be extended towards the reference lists of journal articles. The innovation of *citations indexing* (Garfield, 1979) makes use of internal links between documents, created by the authors themselves. Prior to becoming the playground for science metrics, the *Science Citation Index* was and is (now as part of the *Web of Knowledge*) a very effective tool for information retrieval. Originally distributed on paper, the *Science Citation Index* successfully made the transition to the web. A web-born innovation to select, order and curate knowledge universally is *Wikipedia*. Initiated and hosted by the Wikimedia Foundation, *Wikipedia* is a collective, self-organized encyclopaedia enterprise; driving on user-generated content, web-technologies and principles of free knowledge [8] (Lovink et al. 2011, Dijck 2013). Since 2004 *Wikipedia* operates with a category system - a folksonomy, which increasingly takes a shape similar to classification systems as the UDC (Suchecki et al. 2012).

All of those bibliographical tools have been created to navigate through and to manage growing information spaces. All of them make use of Knowledge Organization Systems [9] (KOS) means to order knowledge and to enable easier and faster access. How to use them is still high on the agenda of information professionals in curricula about information literacy. [10] KOS were the backbone of the organization of collections in physical libraries. The reader might remember the specifically designed furniture at the high time of card-based catalogues. Coming with a search term, one would pull out a drawer and could immediately see: is this a term which is used in the subject catalogue? If so, what items can be found under this term? Are those books nearly what the seeker would expect to find? From which period in time they would be? What other books, what other terms would catch the eye?

## 4. Why it is still more exciting to visit a library than to 'visit' an OPAC of its collection?

With the introduction of Open Public Access Catalogues (OPAC) and aggregators as WorldCat and Europeana, the world's libraries can be visited remotely from behind your computer. While this is an amazing achievement, OPAC's were and are still not easy to use (Borgman, 1996). Based on KOS, the efforts in Information Retrieval seemed to have been focussed on paving the search path of the user, leading her or him seamlessly to the 'right' hits. The Knowledge Organization Systems themselves become hidden in the amazing and complex search engine, which operates behind a search-string-based interface the user is confined to. As Bron (2013, ch2, p. 32) points out: "The familiar design of web search interfaces allows the user to input a keyword-based query via a single search box and presents results as a ranked list of snippets. This type of interface is optimized for looking up facts. A user's information seeking process, however, does not necessarily stop after submitting a query and inspecting some results." Long lists of results make it hard to gain an overview about the available body of literature. In the best case, the user is provided with plain numbers (a kind of statistics) behind the facets [11] of a search, *wordels* or an interactive network visualization of related terms (Figure 2).

What we loose in on-line searching is the tangibility, the physical nature of the card catalogue and the book shelf, which transports effortless information about size and composition of a collection and which allows to browse without losing sight of the whole of the collection (or its part) in which we search. The role of serendipity – the pleasant surprise or, as a physicist would call it, the role of fluctuations – for successful searching has been much discussed among information scientists in the context of scrolling along the shelves of an open stack library. It equally holds for scrolling along the cards - the metadata (Buckland, 1992). An easy switch between the close inspection and a feeling for the whole is what we are after when calling for knowledge maps for collection.

The early architects of large (bibliographic) information space were very sensitive also to visual representations used for information navigation. This holds in particular for Paul Otlet, who designed a *Visual Encyclopedia* as part of the Mundaneum (Van den Heuvel and Rayward, 2011; Kouw et al., 2013). The first *Atlas of Science* was based on products of the ISI and contained network visualizations of research fields and research fronts. (Garfield, 1982) Wikipedia does not operate with overview maps, but there are many visualization and visual browsing experiments based on Wikipedia data. Static, large scale visualization on *Wikipedia* have been produced by Katy Börner's lab (see http://scimaps.org/web/maps/wikipedia/ ). More recently Apps such as LearnDiscovery [12] or WikiWeb [13] offer a visual enhanced navigation through *Wikipedia*.

The question of suitable visual aids for navigation has never left Information Retrieval (Shneiderman, 1994; Börner and Chen, 2002; Shen et al., 2003). However, despite the proclaimed visual turn in society, fostered by tablets and smartphones, visual aids for searching, browsing and navigation come still in the form of experiments or tailored tools for specific communities (cf. Tangherlini, 2013). Although there are more and more original explorations of visual browsing [14], so that it would be worthwhile to review those explorations systematically, the problem remains that the 'usual' digital interface to a library, archive or museum collection has not changed.

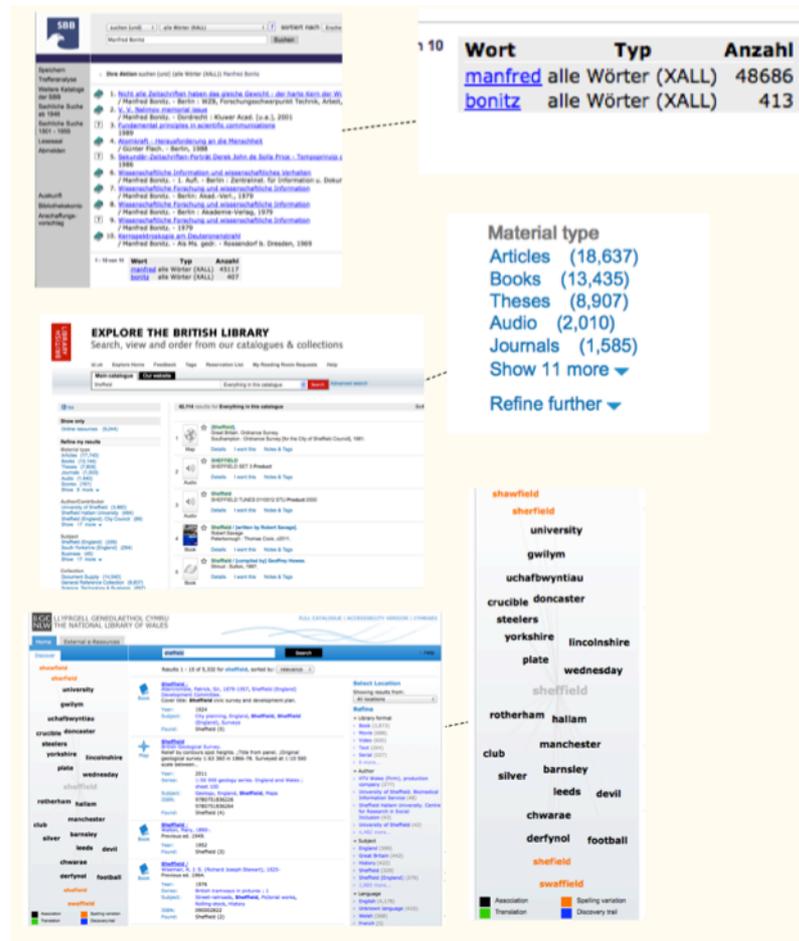

Figure 2: Snapshots of interfaces to the catalogue of the Staatsbibliothek Preussischer Kulturbesitz Berlin (search "Manfred Bonitz", British Library and the National Library of Wales (search "Sheffield") (left top-down) and enlarged parts of the hit lists (right). Only the National Library of Wales offers with the AquaBrowser a visual navigation aid.

## 5. Looking over fences – the *Places&Spaces* Project

When asking Information Retrieval experts about the apparent lack of visual browsing tools, they point to technological barriers. Map-like interfaces to collections require a) to create a (or several) visual representation of a collection which might contain millions of objects and b) allow to click through maps; all high-computing tasks. From the side of visualization it is only recently, that information visualization of large-scale data has left the corner of highly specialized expertise, craftsmanship and tacit knowledge next to access to specific computer resources. Projects as *ManyEyes* [15] and open source libraries as D3.js [16] contribute to a democratisation and dissemination of *good* visualizations.

Another reason might be that librarians and archivists are not the only authorities anymore when it comes to collection, curation and provision of access to knowledge. To some extent "we all are information scientists" [17]. Information sciences, knowledge management, quantitative and qualitative studies of science, behavioural sciences, computer sciences, artificial intelligence, information visualization, infographics and design, educational sciences and science policy are all lined up to propose concepts, models, theoretical studies and practical solutions to the problem of finding the right information at the right place and at the right moment. At the same time, this lengthy summation points to a problem. The potential map makers suffer from the same syndrome as many other researchers: they are many, they operate independently from each other, their knowledge often does not relate to each other. They would need a knowledge map themselves.

The emergence of so-called *science maps*, large scale visual representations based on traces of scholarly communication (journal-based) provides an interesting example of how visualization techniques successfully diffuse into other areas of science.  The first *science maps* appeared at scientometric conferences around the mid 2000ths. Back then, Richard Klavans presented a map, which showed *all* scientific disciplines covered in bibliographic databases, including the social sciences and humanities (Boyack et al., 2005). After this wake-up call, the existing branch of bibliometric mapping in scientometrics (cf. Noyons, 1999) gained momentum. Supported by developments in popular network analysis tools as *Pajek* (Nooy et al., 2005) and new tool developments, scientometrics now uses a couple of different techniques and tools to produce large-scale maps of science. [18] Porter, Rafols and Leydesdorff promoted an overlay technique, which allows displaying activities of institutions, individual researchers or communities over a general map of science (Rafols et al., 2010). Overlay techniques, applied to different base maps of science, have proven to be very popular in use. The *UCSD Map* (Börner, 2012), for instance, has been used in network tools, as Sci[2] (Sci[2] team, 2009), and in semantic web applications as VIVO [19].

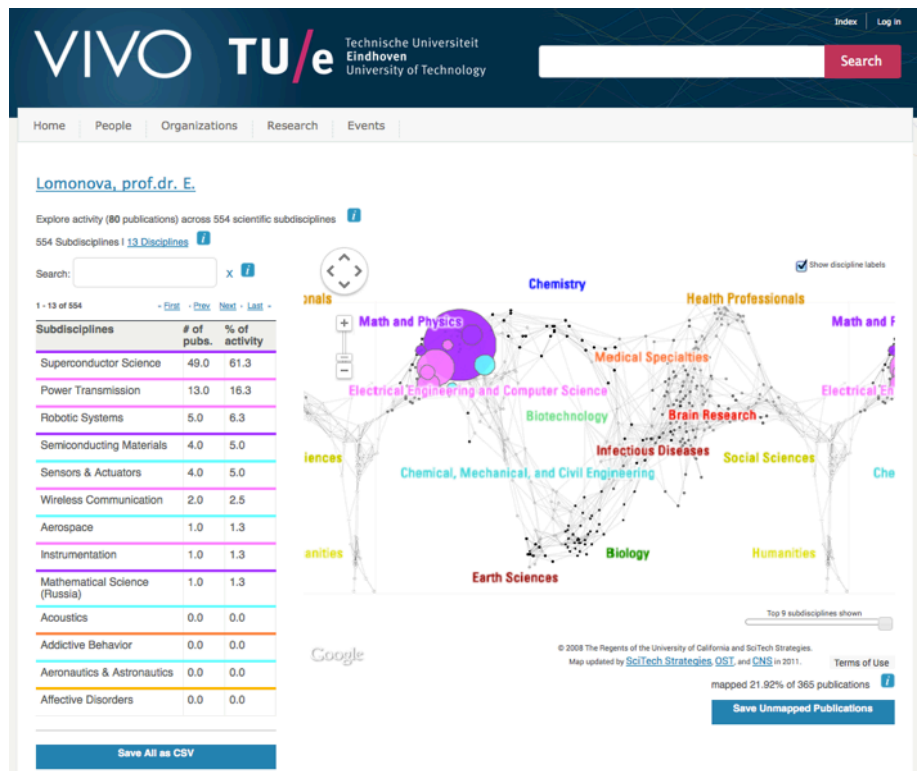

Figure 3 Locations of the publications of Professor Elena Lomonova at a map of science. Prof. Lomonova holds the Chair of Electromechanics and Power Electronics at the Technical University Eindhoven, The Netherlands. The interactive design uses VIVO and has been built by Nick Veenstra and colleagues from the TU/e library. Source: http://vivo.libr.tue.nl/individual?uri=http%3A%2F%2Fvivo.libr.tue.nl%2Findividual%2Fstaff20001731

Many science maps have been featured in the exhibition *Places and Spaces (P&S)*, a ten-year NSF grant effort "to best track and communicate human activity and scientific progress on a global scale" (quote from http://www.scimaps.org/ ). The main curator of *P&S*, Katy Börner, has a computer engineer background as well as joined scientometrics and the information sciences. *P&S* contains more than science maps around scholarly communication. The themes for the yearly editions (called *iterations*) reach from *Science Maps for Economic Decision Makers* to *Science Maps for Kids*. In general, *P&S* has grown to a collection of maps produced by a distributed virtual lab, which included map-makers from various disciplines.

There are different lessons to be learned from *P&S*, when searching for maps for digital collections. First, map making is a long-term activity – think in terms of the history of our geographic maps (Wilford, 2001). Some of the map-making principles are generic, and some of the old maps are still very inspirational. Consequently, *iterations* of *P&S* start with historic maps. Second, map making is a business, which requires many different skills. *P&S* operated from the beginning with open calls for the submission of maps. Criteria for acceptance were the match with the call theme, and the originality of data, but not per se already high-end map-maker skills. *P&S* offers help with the final map making and so contributes to the democratisation process of visualizations. Third, map-making in a vivid technological environment requires continuous technological scouting and brainstorming. Next to the exhibit, Katy Börner runs a sequence of workshop with topics from "Knowledge Management and Visualization Tools in Support of Discovery" to "Exploiting Big Data Semantics for Translational Medicine". Forth, a badly designed map is a useless map. Design and aesthetic principles are inherent to map making. None of the accepted maps for an *iteration* escapes a severe redesign process. Having a designer on board of the *P&S* team was not only useful for dissemination and branding of the project itself, but indispensable to guarantee maps of a certain quality. Fifth, map making requires documentation, and each map needs a legend. To fully understand a map and to foster the diffusion of map-making, one needs to describe the process of the map making as well as its final results. The Atlas of Science by Katy Börner (2010) is an exhibition catalogue, a compendium for science maps, and a comprehensive introduction in map-making techniques. But, reading about maps can only be a first step into map making. At the end there is no escape from the actual practice. So, sixth, a course has been designed as *MOOC* – massive open online course – around principles of visualization and map making (Börner and Polley, 2014). Last but not least, what *P&S* demonstrates and what Katy Börner emphasises in her lectures: there is no good visualization without good data. Seventh, looking for interesting data becomes part of the map making. Concerning science maps, there is a lot of data around the products of science (publications) but the input into science – human capital, instruments, grants – are still largely uncovered ground in terms of standardized information. The so-called research information systems are designed locally (Dijk and Van Meel, 2010), at best at a national level. Semantic web technologies and standardization of ontologies open a window into a larger open, linked, and distributed information spaces, for science as well as for other areas of human activity. Consequently, Katy Börner got involved in VIVO, another US funded project, now on the side of data modelling and data aggregation. In the newly emerging international data space for research information (Börner et al., 2012), maps as the *UCSD Map* become needed and useful navigational tools (Figure 3).

Those are the lessons drawn from the experiences of the *P&S* enterprise [20]: reference to history (1); open, interdisciplinary and experimental (2); brainstorm and scouting parallel to the making of maps (3); importance of design (4); role of documentation (5); teaching and education (6); building and contributing to new data spaces (7). They have inspired the setting up of a specific European network – the COST Action [21] TD1210 – to support the cross-disciplinary quest for knowledge maps, turned into interactive interfaces to digital collections with the aim to enhance access to Europe's cultural heritage. [22]

# 6. KnoweScape – a newly formed coalition of knowledge map makers

At the beginning of *KnoweScape* stands a map *Design versus Emergence* (or in short Wikipedia-UDC map). This map displays the category system of Wikipedia next to the tree structure of the classes in the UDC (Figure 4). The Wikipedia categories grow bottom up in a self-organized, editing process from million actions of tagging Wikipedia articles with uncontrolled vocabulary (Suchecki et al., 2012). The UDC is a system to express concepts, designed by Paul Otlet and developed and revised since the beginning by a small circle of devoted editors. Both can be visually compared applying a shared colour scheme based on the careful alignment of the different term sets (Akdag et al., 2011a). What if one could use such a map to surf through Wikipedia or to surf through the catalogue of a National Library? What if lights would flicker on this map whenever a search term is entered to indicate where in this information universe the search landed?

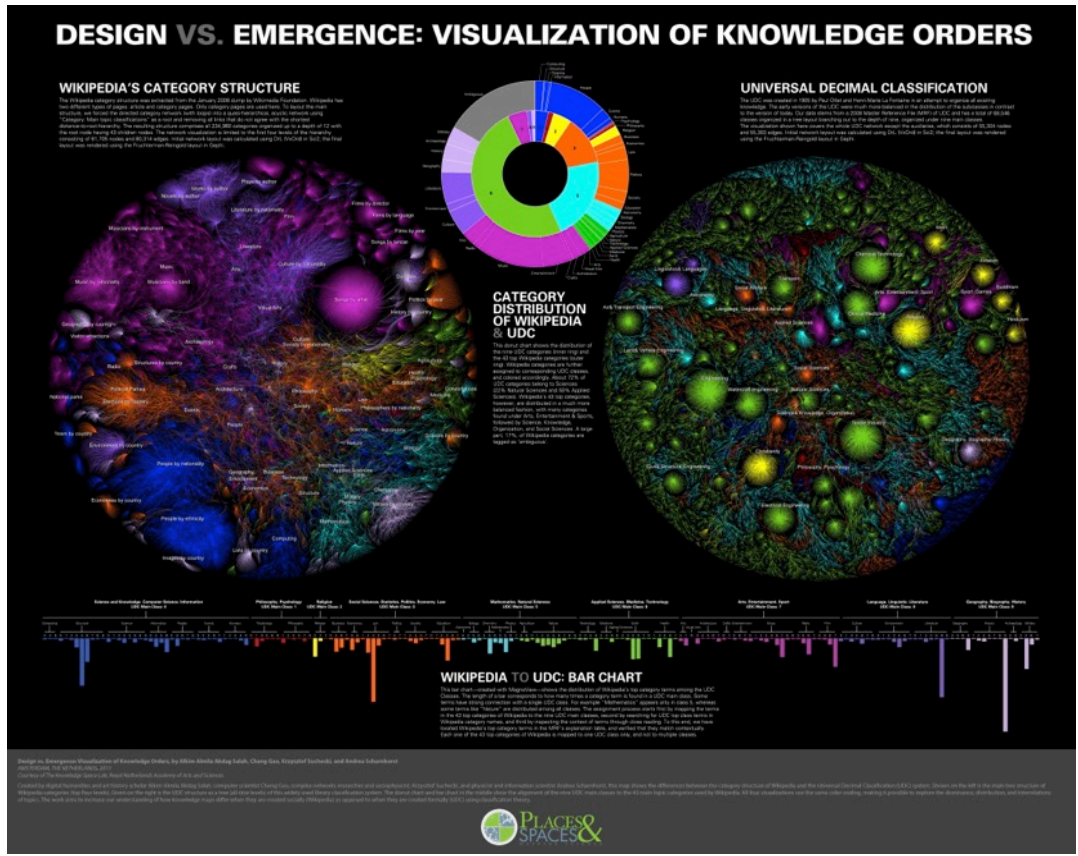

Figure 4 Design vs. Emergence: Visualisation of Knowledge Orders. (Akdag et al. 2011b)

The Wikipedia-UDC map was one of the results of the *Knowledge Space Lab*, a project funded by the Royal Netherlands Academy of Arts and Sciences, and hosted by the *Virtual Knowledge Studio* (Wouters et al., 2013). The project team consisted of a computer scientist, an art historian/digital humanities scholar, a physicist and an information scientist. At the time of the project, a COST Action MP0801 "Physics of competition and conflicts" (Richmond, 2012) formed the international home for this project, and many members of this COST Action contributed also to TD1210 – *KnoweScape. KnoweScape* is clearly devoted to an information science problem, and one might wonder what this has to do with physics?

In the last decades, methods from statistical physics have been applied to social, economic and cultural processes, leading to fields as sociophysics (cf. Galam, 2012) and econophysics (cf. Schweitzer, 2007). Those are tendrils or interfaces from physics into social sciences and humanities, but the core research community behind them is still part of physics, at least for what concerns epistemics, concepts, and methods. Statistical physics deals with many particle systems, and the power of its epistemics unfolds when it comes to pattern (structure) recognition in large data spaces, and the search for dynamic processes behind their evolution. This kind of physics is part of the wider community of complexity research. However, as original and surprising the detected patterns might be, they have shortcomings, which could be labelled as being "orphan structures" or "structures without names". Whenever it comes to labelling patterns and structures found in on-line listening behaviour, co-authoring text, or co-occurrence of names from a political elite, one needs context information to really harvest knowledge from newly found patterns.

As an illustration: making the Wikipedia map required a reconstruction of the evolution of links between article pages and category pages from the 2.8 TB large dump from Wikimedia. This was done with grid computing [23]. But, the interpretation of extracted changes profited or even required the comparison to other knowledge organization systems, such as the UDC.

This is the reason for the current composition of communities in *KnoweScape* [24] (Figure 5). For the first time, a platform has been created where information professionals, sociologists, physicists, digital humanities scholars and computer scientists collaborate on problems of data mining and data curation in collections. Traces of human knowledge production, preserved by information specialists (librarians, archivists, scholars in humanities fields, museum curators, contributors of Web spaces) provide the empirical base of research in this Action. The methodological basis for pattern recognition comes from complexity and network science. The toolbox of statistical physics concepts and methods forms the technical core. New insights resulting from evolutionary, dynamic approach to knowledge systems will be combined with new methods to index, order, and retrieve digital information from large knowledge spaces. In the analysis of knowledge ordering systems and data from large collections, mathematical models of collective knowledge production as developed in computational sociology, computational philosophy and socio-physics will be used. New navigation tools will be developed in collaboration with knowledge-domain specialists in the social sciences and humanities. Semantic web approaches will be used to enhance access to collections and to also link between their different knowledge organizations.

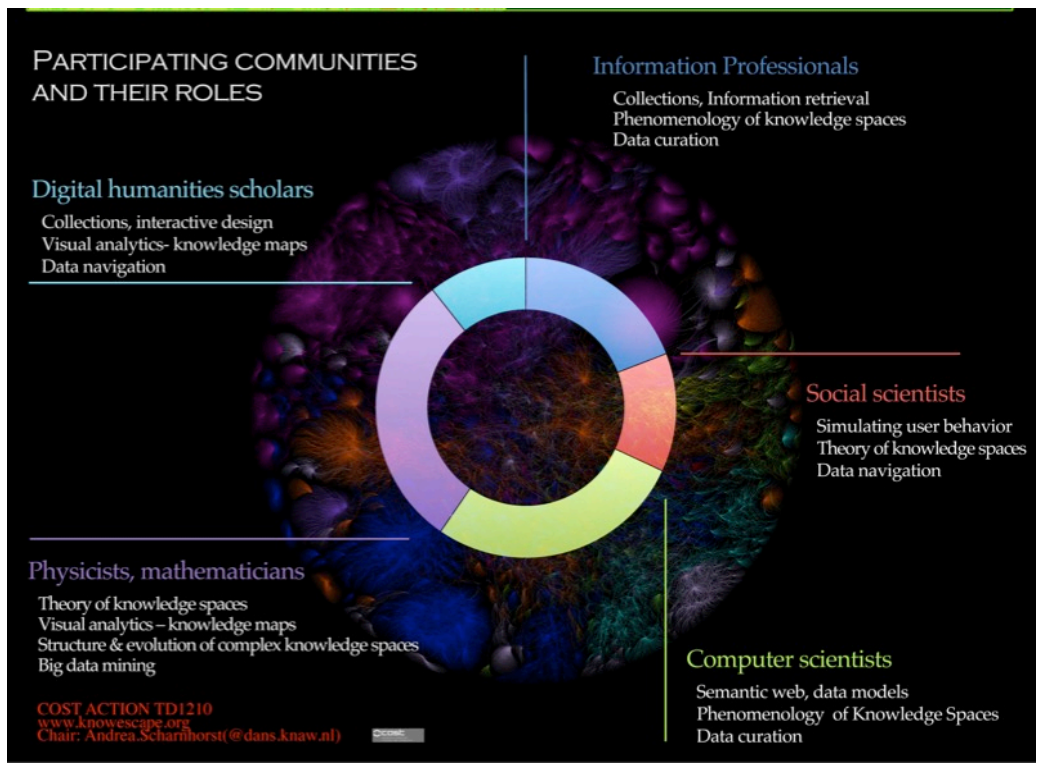

In four different working groups, researchers in KnoweScape work on the following five tasks.

(1) *Selection of knowledge spaces to be mapped*: The notion of a "knowledge space" is used in a broad sense. Empirically, we rely on collections of artefacts (and their description) as studied in the humanities; library, archival and museum collections and national bibliographies; but also web-based information spaces such as Wikipedia, social networks (e.g. Facebook), and collaborative platforms (e.g. Mendeley, CiteUlike). Scholarly communication as recorded in commercial abstracting and indexing bibliographic databases such as Current Contents, ERIC, Medline, SPIN, CAS etc. available through service providers such as ISI Web of Science, Scopus, PubMed; national research information systems; and databases of funding institutions is one area where the interests of different communities overlap.

(2) *Data representations to support data mining across knowledge spaces*: The fact that knowledge spaces are available on-line does not automatically imply that information about them can be easily harvested. Classification systems are usually designed for the information needs of specific communities of users. How to map them in order to create links between different sources of knowledge has been addressed extensively in vocabulary mapping projects such as Multilingual Access to Subjects (MACS), Renardus, Semantic Interoperability to Access Cultural Heritage (STICHT), Multilingual Subject Access to Catalogues (MSAC) and terminology services projects such as OCLC Metadata Switch and High Level Thesaurus Project (HILT). Recently, the W3C linked data activities within the Semantic Web initiative (a continuation of the SKOS standard development), and promoted publishing and sharing of knowledge organization systems on the web as well as their mapping. While under pressure to organize the ever-increasing amount of digital information there is a danger of overlooking data, information, and knowledge that have now become part of our scientific and cultural heritage. Combining traditional and recent ways of knowledge organization, this network addresses the question of how to define an appropriate generic level on which knowledge organization systems can be linked.

(3) *Identify dimensions of knowledge maps – modelling and analysing knowledge spaces*: To design a knowledge map requires the identification of characteristic attributes and forms of representation. Knowledge maps can take very different forms, and purposefully, we don't start with a fixed definition. Examples are treemaps representing the disciplinary composition of a collection. Timelines as used in Europeana Connect can be combined with displaying the geographic spread of objects in a collection. But, there might be alternative reference systems (similar to a parameter or variable space). To find those hidden dimensions useful to be represented on a map requires pattern extraction based on similarities of knowledge objects (e.g., book, article, author, text piece, institution) as expressed on their description (text), authorship, form and media, age of production and other metadata. Usual top-down approaches in knowledge organization need to be combined with bottom-up approaches to mining relational information among knowledge objects. *KnoweScape* takes the problems of knowledge organization back to its roots by applying a "non-linear physics" (or complexity) inspired perspective. Collections are seen as large-scale systems composed from knowledge objects. Knowledge ordering systems are seen as attributes or coordinates. In the combination of both, we seek for structures and their evolution in time.

Empirical analysis is one part of a better understanding of complex knowledge spaces. The application of dynamic models to explain their emergence and how to navigate through them is another important part of the theoretical work to be done. Hereby, combining models from physics and computational sociology can help to simulate artificial knowledge spaces and to test the use of knowledge maps for human information navigation.

(4) *A typology of knowledge maps*: Ideally, for one collection different knowledge maps would be available between which a user would be able to choose and switch. Another feature could be knowledge maps which adapt to the search behaviour of the user. To be able to implement such features, one first needs to have a variety of different representations to choose from. In the case of large amounts of information, algorithms have to be developed that allow the display of information on the fly and on the web. KnoweScape produces a catalogue of different knowledge maps including requirements for input data format, computational requirements and tools, and possible uses based on insights and experiences of scientific visualization specialists. It also collects sketches for knowledge maps produced in brainstorm sessions.

(5) *Implementation of knowledge maps*: It would go beyond the capacity of a COST Action to already implement knowledge maps in living on-line interfaces. To integrate new interfaces into living services is a complicated, multi-step process in itself. But, KnoweScape supports research in this area and aims to develop 'proof of concept' demonstrators. By documenting the research process and by communicating closely with information professionals and other stakeholders, KnoweScape develops workflows and guidelines suitable for future implementation of knowledge maps.

## 6. Conclusion: Dream your library – dream your knowledge map [25]

In this paper we started from changes in the world of digital scholarship, and drew the conclusion that sometimes maps might be helpful to master complex information spaces. We looked briefly into the history of knowledge organization and information science. We pointed out the need to form a broad alliance in the quest for knowledge maps as interfaces to digital collections. We proposed an alliance, which spans a bridge from librarianship to physics. We looked into *Places and Spaces* as a possible role model to organize the quest for knowledge maps, and laid out the principles around the European collaborative network *KnoweScape* – a COST Action just started.

While knowledge maps have not yet left the phase of tinkering in workshops of potential map makers, we can at least dream about knowledge maps and envision scenarios for their use. If we one more time return to the history of bibliometric mapping, there are two lessons to be learned: first, topic maps for specialities can best be interpreted by specialists, and second, those specialists most of the time have a fairly good internal map of their territory and actually don't need maps. But there are other, more mundane areas of knowledge acquisition, which would profit from a concerted action of new ways of knowledge representation, visual analytics and interactive interfaces to knowledge management. Let us as a conclusion design three such use cases.

Imagine an organization that aims to start a course on data management and data stewardship. This could be a Trusted Digital Repository or a funding agency, but could also be a university library. There are syllabus and resources globally available which address these issues, all in in their own way. There are professional organizations, as the newly formed Research Data Alliance, with working groups, manifestos, wiki's. There are recommendations, policy documents, good practices on data sharing and Open Access policies. There are 30 years or more of studies in this area. A wealth of information and experiences world-wide from which course material can be composed by the concrete usual small team of people at that imagined organization. Would the course designers not be helped by a map, which shows them what is already available? Made by which institution? A geographic map of course material, color-coded according to the disciplinary perspective, with timelines and names of experts available as lecturers? This would allow to cherry-pick elements most suitable for the local need of a course in data management. It would also ensure that this course is up to scratch. It would lift individual, social network-based and web-search based information foraging to a more effective level. Eventually, the concrete course, implemented and adapted locally their course material would be feed back – flag out on the map, re-usable by others, so that they become part of a wider pool of training resources.

Imagine a science policy maker who sets up a programme in, say, Digital Humanities. Imagine this decision maker having at hand a continuous, in-time, up-to-date monitoring of investments made in the past, on-programs world-wide and a investment versus return analysis. The usual practice is continuous reporting and web-based communication from the side of the grant-takers. Current practice is also to set up expert-based in-depth briefings and evaluative reports at the beginning and at the end of a funding stream. In some cases, infographics or a map are commissioned based on scientific information visualization for which data are collected, curated and cleaned. The result is a case study and a static visual snapshot. What about a large-scale display wall on which you could inspect the growth and diffusion of a certain scientific activity by streaming twitter data, upcoming conferences, experts in the fields, establishing of chairs at universities world-wide, patents and emergence of SME related to this new field?

Imagine a student confronted with a question and seeking where best methods and concepts for this question can be found. Imagine a researcher obliged to start a research collaboration with colleagues when their academic language, way of reasoning, norms and history of the field are unknown to her or known only in very general terms. Would those not be supported by a dashboard which gives them: milestones in history and famous names (from Wikipedia), current leading centers (from Vivoweb), most used notions (from blogs, Mendeley tags and delicious bookmarks), highly cited papers and shortest path between their own collaborative network and those of the colleagues with whom there will be collaboration in the future (as in Microsoft Academic Search)? A dashboard which is interactive, working on mobile devices, which returns search terms with graphics and not 'just' a list of hits. Something which acts as a supportive knowledge agent – very similar to the librarian or documentalist one can approach with a question; in other words an Information Specialist App.

A precondition for analysis and visual representations as envisioned in the scenarios above is the existence of a data representation of different information spaces involved, which can be read by machines and recombined as needed. In other words, something like the Linked Open Data cloud are needed. The well-curated content from library and archive collections should be part of it, indispensible if we want to build on the historical acquired knowledge.

But in order to enable machines to bring the scattered knowledge together and to visualize it, translatory work among the different communities in possession of knowledge and skills about knowledge organization, management and analysis is needed. This paper described the circumstances, under which old dreams of knowledge maps get revitalized, takes a glance at the current state of knowledge maps in theory and practice, and eventually introduces one specific research network KnoweScape as a means among others to make them.

### Acknowledgement:

This work has been funded by the COST Action TD1210 KnoweScape. Part of section 5 bases on the Memorandum of Understanding of TD1210. Special thanks goes to Aida Slavic, Peter Mutschke, Peter Richmond, Giulia Rotundo, Marcel

Ausloos, Bruce Edmonds and Nigel Gilbert contributed extensively to the MoU. Part of the works has also been funded by EINS – a Network of Excellence for Internet Science. Thanks for comments to anonymous reviewers, and Katy Börner and Christophe Gueret who commented on an earlier version of the text.

---

## Endnotes:

[1] www.knowescape.org  See also its Memorandum of Understanding.
[2] http://en.wikipedia.org/wiki/Association_of_Internet_Researchers
[3] Throughout this paper the term *science* is used to address all scholarship and academia. Rather than its modern use in English for « systematic study of the structure and behaviour of the physical and natural world through observation and experiment: the world of science and technology» (Dictionary, Apple 2.2.3 (118.5)) we use the term as one would use *Wissenschaft* in German, *les sciences* in French, *wetenschap* in Dutch or наука in Russian.
[4] http://en.wikiquote.org/wiki/E._O._Wilson
[5] The so-called *Deltawerken* (or Delta Works) are a complex system of  dams, sluices, locks, dykes, levees, and storm surge barriers to protect the highly populated area in the river delta formed by  Rhine, Meuse and Scheldt. (see http://en.wikipedia.org/wiki/Delta_Works  (Bijker, 1995).

[6] In 2012 the Mundaneum organized an exhibition "Renaissance 2.0" with a comprehensive view on early classification of knowledge up to the Internet age. Quotation from the website "This new area of distribution of knowledge has undergone premises and has formed thanks to the work of pioneers who succeeded in tracing a clear vision. Paul Otlet and Henri La Fontaine, founders of the Mundaneum, are amongst those visionaries and their task is nothing short of an edifying testimony. These 'Classifiers' of the world and those who preceded them (the scholars of the Middle Ages to the collectors princes of the Renaissance and the encyclopaedists of Enlightenments) tried to give sense to the empirical advance of knowledge and offer a clear filing method to facilitate its access." (see http://expositions.mundaneum.org/en/renaissance-20-read-more )

[7] The principles of indexing applied by Google and the Repertoire Bibliographic Universel are very different, namely keywords-based versus taxonomy-based. What is similar is that user can send in their questions and get back answers.

[8] https://meta.wikimedia.org/wiki/Wikimedia_Foundation/Annual_Report/2012-13

[9] « Knowledge Organization Systems (or KOS) is a generic term used in Knowledge organization about authority lists, classification systems, thesauri, topic maps, ontologies etc. » Source : Wikipedia, http://en.wikipedia.org/wiki/Knowledge_Organization_Systems

[10] See http://www.ala.org/acrl/standards/informationliteracycompetency

[11] Facets is another technical term from the information sciences. Facets are attributes which are used next to classes to order objects. In the UDC classes « Main Tables » represent fields or disciplines containing concepts used for the study of phenomena. For instance, Table 3 stands for social sciences. Additionally, the UDC uses auxiliaries (language, form, place, time, ethnicity, properties). Those are generic, and systems of ordering orthogonal to the topical, hierarchical organization. With facets present one can for instance ask: « give me all books about sociology in Hungarian language in the 20th century ». Hereby, sociology is a class, but book, Hungarian language, and 20th century, are facets (see also McIlwaine, 2010).

[12] http://learn.xdiscovery.com/

[13] http://www.wikiwebapp.com/

[14] Sometimes they are called Graphical Knowledge Engine. See Eyeplorer as an example http://en.vionto.com/show/me/eyePlorer.com?login=1 . See the PATH project http://www.paths-project.eu/eng , see also the last UDC conference http://seminar.udcc.org/2013/programme.php   (Slavic et al., 2013)

[15] http://www-958.ibm.com/software/data/cognos/manyeyes/

[16] http://d3js.org/  The website used a kind of visual navigation aid.

[17] This was the answer given by Martin White in his keynote talk at the First Annual KnoweScape Conference, when asking the audience « who is an information scientist ? » See : http://knowescape.org/knowescape2013/

[18] Examples are the tools VosViewer and CitNetExplorer by the CWTS, http://www.vosviewer.com/ and http://www.citnetexplorer.nl/Home (Van Eck and Waltman, 2007); the overlaykit tool set from Loet Leydesdorff website: http://www.leydesdorff.net/journalmaps/ ; the Sci2 tool of Katy Börner's lab